# Quantum Mind from a Classical Field Theory of the Brain

P. Zizzi
Department of Psychology, University of Pavia,
Piazza Botta, 6, 27100 Pavia, Italy
paola.zizzi@unipv.it

**Abstract**
We suggest that, with regard to a theory of quantum mind, brain processes can be described by a classical, dissipative, non-abelian gauge theory. In fact, such a theory has a hidden quantum nature due to its non-abelian character, which is revealed through dissipation, when the theory reduces to a quantum vacuum, where temperatures are of the order of absolute zero, and coherence of quantum states is preserved. We consider in particular the case of pure SU(2) gauge theory with a special anzatz for the gauge field, which breaks Lorentz invariance. In the ansatz, a contraction mapping plays the role of dissipation. In the limit of maximal dissipation, which corresponds to the attractive fixed point of the contraction mapping, the gauge fields reduce, up to constant factors, to the Pauli quantum gates for one-qubit states. Then tubuline-qubits can be processed in the quantum vacuum of the classical field theory of the brain, where decoherence is avoided due to the extremely low temperature. Finally, we interpret the classical SU(2) dissipative gauge theory as the quantum metalanguage (relative to the quantum logic of qubits), which holds the non-algorithmic aspect of the mind.

# 1. Introduction

Hameroff and Penrose suggested, in their orch. OR model [1] [2] of the Quantum Mind, that tubulines in microtubules can be in superposed states, like qubits, leading to quantum computation in the brain.

Influential criticism of the possibility that quantum states can in fact survive long enough in the thermal environment of the brain has been raised by Tegmark [3]. He estimates the decoherence time of tubulin superpositions due to interactions in the brain to be less than $10^{-12}$ sec. Compared to typical time scales of microtubule processes of the order of milliseconds and more, he concludes that the lifetime of tubulin superpositions is much too short to be significant for neurophysiological processes in the microtubule.

In a response to this criticism, Hagan *et al.* [4] have shown that a revised version of Tegmark's model provides decoherence times up to 10 to 100 μ sec, and it has been argued that this can be extended up to the neurophysiologically relevant range of 10 to 100 msec under particular assumptions of the scenario by Penrose and Hameroff.

In this paper, we suggest that tubulines-qubits can be processed in the quantum vacuum (where temperatures are of the order of absolute zero, and coherence is maintained) of a classical dissipative non-abelian gauge theory of the brain.

In a very recent paper [5] we considered the particular case of a classical SU(2) Yang-Mills theory. Such a theory has an hidden quantum nature, due to its non-abelian character. In fact, it exhibits a quantum vacuum if dissipation is taken into account. In [5], the role of dissipation was played by a contraction mapping in a particular ansatz for the gauge field, which breaks Lorentz invariance. In a limit of the ansatz corresponding to the attractor, the theory falls in a quantum vacuum. There, the gauge field components reduce to quantum logic gates of one-qubit states.

The idea of describing brain processes in terms of a field theory goes back to the1960s, when Ricciardi and Umezawa [6] suggested to utilize the formalism of quantum field theory to describe brain states, with particular emphasis on memory (the "Quantum Brain Dynamics" paradigm).

In Quantum Brain Dynamics, the field theory is quantum from the start, while in this paper we consider a classical field theory, and look for its hidden quantum features.

The proposal of Ricciardi and Umezawa has gone through several refinements, for example by Stuart and Major [7] and by Jibu and Yasue [8]. Some more recent progress has been achieved by Vitiello [9] by including dissipation. However, making dissipation to agree with quantization is an hard task, due to the appearance of non- Hermitian operators, and in fact Vitiello's dissipative quantum field theory encountered some technical difficulties. More precisely, the concrete building of a Dissipative Quantum Field Theory requires a generalization of the usual Quantum Field Theory. Namely the latter is based on assemblies of harmonic oscillators, which, in the case of dissipative processes, should be replaced by *damped oscillators*. Unfortunately the latter do not fulfil the energy conservation principle, and this fact makes unreliable any attempt to introducing an Hamiltonian-like formalism. A convenient strategy was introduced in [10] [11]where it was described the influence of a dissipating environment by *doubling* the original damped system through the introduction of a time-reversed version of it, which acts as an absorber of the energy dissipated by the original system.

More recently, Vitiello and collaborators [12] presented an example of dissipation in a classical system which explicitly leads, under suitable conditions, to a quantum behaviour. They showed that the dissipation term in the Hamiltonian for a couple of classical damped-amplified oscillators manifests itself as a geometric phase and is actually responsible for the appearance of the zero point energy in the quantum spectrum of the 1D linear harmonic oscillator. It seems that the our and their lines of thought have some point in common and are fundamentally in agreement. Some of the assumptions in [12] were inspired by 't Hooft work [13] [14], where he discussed classical, deterministic, dissipative models and showed that constraints imposed on the solutions which introduce information loss resembles a quantum structure. 't Hooft's conjecture is that the dissipation of information which would occur at Planck scale in a regime of completely deterministic dynamics would play a role in the quantum mechanical nature of our world.

Penrose's idea of the non algorithmic nature of mathematical intuition [15] [16] is another important feature of his vision of the Quantum Mind. Here we support this idea, although we use quantum metalanguage [17] instead of the first Gödel's incompleteness theorem.

However, the two approaches are related to each other, once one takes into account the quantum version [18] of Gödel's theorem, in the logic of quantum information, which derives from a quantum metalanguage.

The paper is organized as follows:

In Sect. 2, we present the physical model, that is, the classical SU(2) gauge theory, the ansatz, which breaks Lorentz invariance, and the contraction mapping playing the role of dissipation.

In Sect. 3, we show that qubits can be processed in the quantum vacuum due to the fact that there the gauge field components reduce to quantum logic gates for one-qubit states. Due to the very low temperature of the quantum vacuum, tubuline-qubits do not decohere.

In Sect. 4, we interpret the classical dissipative non-abelian gauge theory of the brain as the quantum metalanguage, from which originates the quantum object language of the unconscious, and argue that quantum metalanguage represents the non-algorithmic aspect of the mind.

## 2. The Physical Model

There is an ansatz [5] for the classical SU(2) gauge field, which, in a particular limit corresponding to a vacuum solution, enables one to recover spin ½ quantum mechanics. This ansatz is gauge invariant, but breaks Lorentz invariance. Of course the nature of the new vacuum state must be intrinsically quantum. At this point one might ask which is the physical mechanism that can trigger this process, which leads to a quantum vacuum state of the original classical theory. The most plausible answer is dissipation. A dissipative system is characterized by the spontaneous appearance of symmetry breaking, which in our case is the breaking of Lorentz symmetry. This vacuum is quantum as all the thermal fluctuations have disappeared because of dissipation, and quantum fluctuations dominate. In the limit, the gauge field reduces to the generator of a global U(1), i.e., a phase, times a Pauli matrix, that is, a quantum logic gate of one-qubit state. This suggests that qubits can be processed in a quantum vacuum of the classical SU(2) gauge theory. Given the quantum vacuum is at zero absolute temperature, $T = 0$, the qubits do not decohere, unless they are put in interaction with an external environment. In [5] we did not describe dissipation by any particular model, however the role of dissipation was played by a contraction mapping in the ansatz. The contraction mapping is related to some geometrical aspects of the gauge theory under consideration.

## 2.1. The Ansatz

In [5], we considered the SU(2) gauge field $A_\mu^a(x)$ ($\mu = 0,1,2,3;\ a = 1,2,3$) and made the following ansatz:

$$A_\mu^a(x) = e^{-i\lambda_\mu(x)} \sigma^a \qquad (2.1)$$

where $\lambda_\mu(x)$ is a U(1) gauge field and the $\sigma^a$ are the Pauli matrices, which satisfy the commutation relations:

$$[\sigma^a, \sigma^b] = 2i\varepsilon_{abc}\sigma^c \qquad (2.2)$$

The ansatz (2.1) explicitly breaks Lorentz invariance.

In the following we will consider, in particular, the limit case:

$$\lambda_\mu(x) \to 0 \qquad (2.3)$$

In this limit one gets $A_\mu^a(x) \to \sigma^a$. In a sense, the SU(2) gauge theory reduces to the quantum mechanics of spin ½.

The ansatz (2.1) can be rewritten as:
$$A_\mu = e^{-i\lambda_\mu} \tag{2.4}$$
where:
$$A_\mu \equiv A_\mu^a \sigma^a / 2 \tag{2.5}$$
Let us consider the SU(2) gauge transformations performed on the original gauge field $A_\mu$:
$$A_\mu \xrightarrow{U} A_\mu' = U A_\mu U^{-1} - \frac{i}{g} U \partial_\mu U^{-1} \tag{2.6}$$
where $g$ is the gauge coupling constant, $U$ is given by:
$$U = \exp(i\rho^a(x)\sigma^a / 2) \tag{2.7}$$
and $\rho^a(x)$ are three arbitrary real functions.
The ansatz (2.7) transforms under (2.5) as:
$$e^{-i\lambda_\mu} \xrightarrow{U} e^{-i\lambda_\mu'} = e^{-i\lambda_\mu} - \frac{i}{g} U \partial_\mu U^{-1} \tag{2.8}$$
In the limit case (2.3) the transformations (2.8) become:
$$e^{-i\lambda_\mu} \xrightarrow{U} e^{-i\lambda_\mu'} = 1 - \frac{i}{g} U \partial_\mu U^{-1} \tag{2.9}$$
Eq. (2.9) can be transformed into a pure gauge by a suitable choice of the arbitrary functions $\rho^a(x)$. This means that in the limit case the ansatz (2.1) describes a vacuum solution.

In the original SU(2) theory invariant under Lorentz transformation, the vacuum state was $|0\rangle$, corresponding to $A_\mu = 0$. In presence of the ansatz, which breaks Lorentz invariance, there is, in the limit case, a new vacuum state $|\vartheta\rangle$, corresponding to $A_\mu = 1$.

Then, the gauge field $A_\mu$ has a non-vanishing v.e.v. in the new vacuum:
$$\langle \vartheta | A_\mu | \vartheta \rangle \neq 0 \tag{2.10}$$
Let us take the temporal gauge $A_0 = 0$. Then, Eq. (2.10) becomes:
$$\langle \vartheta | A_i | \vartheta \rangle \neq 0 \qquad (i=1,2,3) \tag{2.11}$$
This indicates that there is a spontaneous symmetry breaking of the little group O(3), to which are associated three Goldstone bosons $\varphi_i$, each one corresponding to a particular O(3) generator.

## 2.2. The contraction mapping as dissipation

The pure SU(2) gauge theory under consideration can be described in terms of a principal fiber bundle $(P, \pi, B, G)$, where $P$ is the total space, $B$ is the base space (in our case $R^4$), $G$ (in our case SU(2)) is the structure group, which is homeomorphic to the fiber space $F$, and $\pi$ is the canonical projection:
$$\pi : P \to R^4 \tag{2.12}$$
(For a review on principal fiber bundles see, for instance Ref. [19]).
The base space $R^4$ is equipped with the Euclidean metric $d$:
$$d(x', x) = |x' - x| \tag{2.13}$$
where $x$ and $x'$ are two points of $R^4$ and must be intended as $x \equiv \{x_\mu\}$, $x' \equiv \{x_\mu'\}$ $(\mu = 1,2,3,4)$.
The complete metric space $(R^4, d)$ has an induced topology which is that of the open balls with rational radii $r_n = \frac{1}{n}$, with $n$ a positive integer.
The open ball of rational radius $r_n$, centred at $x^*$ is:

$$B_{r_n}(x^*) = \{x \in R^4 \mid d(x^*, x) < r_n\} \tag{2.14}$$

The set of open balls $B_{r_n}(x^*)$ is an open covering of $R^4$ and forms a local basis for the topology.

Now, let us consider again the ansatz (2.4), and make the following natural choice for $\lambda_\mu(x)$:

$$\lambda(x) = x^* e^{i\frac{|x^*-x|}{n}} \tag{2.15}$$

where $\lambda$ in (2.15) must be intended as $\lambda \equiv \{\lambda_\mu\}$ $(\mu = 1,2,3,4)$.

The point $x^*$ is a fixed point for $\lambda(x)$ as it holds:

$$\lambda(x^*) = x^* \tag{2.16}$$

It is easy to check that $\lambda(x)$ continuously approaches $x^*$ for large values of $n$:

$$\lim_{n \to \infty} \lambda(x) = x^*. \tag{2.17}$$

The fixed point $x^*$ is an *attractive* fixed point for $\lambda(x)$, as it holds:

$$|\lambda'(x^*)| < 1 \tag{2.18}$$

The point $x^*$ is then a particular kind of attractor for the dynamical system described by this theory. Furthermore, it holds:

$$|\lambda'(x)| < 1 \tag{2.19}$$

for all $x \in B_{r_n}(x^*)$, which is equivalent to say that $\lambda(x)$ is a contraction mapping in the attraction basin of $x^*$, that is, it satisfies the Lipschitz condition [20]:

$$d(\lambda(x), \lambda(x')) \leq q \, d(x, x') \tag{2.20}$$

with $q \in (0,1)$ for every $x, x' \in B_{r_n}(x^*)$.

## 3. Qubits processed in the quantum vacuum

The qubit is the unit of quantum information. It is the quantum analog of the classical bit $\{0,1\}$, with the difference that the qubit can be also in a linear superposition of 0 and 1 at the same time. (For a review on quantum information see, for instance, Ref. [21]).

The qubit is a unit vector in the 2-dimensional complex Hilbert space $C^2$.

The expression of the qubit is:

$$|Q\rangle = \alpha|0\rangle + \beta|1\rangle \tag{3.1}$$

Where the symbol $|\ \rangle$ is the ket vector in the (bra-ket) Dirac notation in the Hilbert space.

The two kets:

$$|0\rangle = \begin{pmatrix} 1 \\ 0 \end{pmatrix}, \quad |1\rangle = \begin{pmatrix} 0 \\ 1 \end{pmatrix} \tag{3.2}$$

form the orthonormal basis of the Hilbert space $C^2$, called the computational basis.

The coefficients $\alpha, \beta$ are complex numbers called probability amplitudes, with the constraint:

$$|\alpha|^2 + |\beta|^2 = 1 \tag{3.3}$$

to make probabilities sum up to one. (Any quantum measurement of the qubit, either gives $|0\rangle$ with probability $|\alpha|^2$, or $|1\rangle$ with probability $|\beta|^2$).

The geometrical representation of the qubit corresponds to the Bloch sphere, which is the sphere $S^2$ with unit radius. Formally, the qubit, which is a point of a two-dimensional vector space with complex coefficients, would have four degrees of freedom, but the constraint (3.3) and the impo ssibility to observe the phase factor reduce the number of degrees of freedom to two.
Then, a qubit can be represented as a point on the surface of a sphere with unit radius.

The Bloch sphere is defined by:

$$S^2 = \left\{ x_i \in R^3 \,\middle|\, \sum_{i=1}^{3} x_i^2 = 1 \right\} \tag{3.4}$$

Any generic 1-qubit state in (3.1) can be rewritten as:

$$|Q\rangle = \cos\frac{\vartheta}{2}|0\rangle + e^{i\phi}\sin\frac{\vartheta}{2}|1\rangle \tag{3.5}$$

where the Euler angles $\vartheta$ and $\phi$ define a point on the unit sphere $S^2$.

Thus, any 1-qubit state can be visualized as a point on the Bloch sphere, the two basis states being the poles.

We remind that any transformation on a qubit during a computational process is a reversible operation, as it is performed by a unitary operator $U$:

$$U^\dagger U = I. \tag{3.6}$$

where $U^\dagger$ is the Hermitian conjugate of $U$.

This can be seen geometrically as follows. Any unitary $2\times 2$ matrix $U$ on the 2-dimensional complex Hilbert space $C^2$ (which is an element of the group SU(2)) multiplied by a global phase factor):

$$U = e^{i\phi}\begin{pmatrix} \alpha & \beta \\ -\beta^* & \alpha^* \end{pmatrix} \tag{3.7}$$

(where $\alpha^*$ is the complex conjugate of $\alpha$), can be rewritten in terms of a rotation of the Bloch sphere:

$$U_2 = e^{i\phi} R_{\hat{n}}(\theta) \tag{3.8}$$

where $R_{\hat{n}}(\theta)$ is the rotation matrix of the Bloch sphere by an angle $\theta$ about an axis $\hat{n}$.

In [5] we showed that the SU(2) gauge fields $A^a_\mu$ reduce to the operators $A^a$ which, up to a multiplicative constant, are the product of the generator of a global U(1) group times the Pauli matrices:

$$A^a = -ige^{-ix^*}\sigma^a \tag{3.9}$$

This means that the pure SU(2) gauge field theory is reduced to a quantum mechanical theory of spin ½ with a constant U(1) "charge", in absence of any interaction.

The operators $A^a$ in (3.9) are unitary operators, as it holds:

$$A^a A^{a\dagger} = 1 \tag{3.10}$$

Then, the $A^a$ operators can play the role of quantum logic gates for one-qubit states. In fact, the X, Y, Z quantum logic gates for one-qubit are just the three Pauli matrices:

$$X = \begin{pmatrix} 0 & 1 \\ 1 & 0 \end{pmatrix} \quad Y = \begin{pmatrix} 0 & -i \\ i & 0 \end{pmatrix} \quad Z = \begin{pmatrix} 1 & 0 \\ 0 & -1 \end{pmatrix} \tag{3.11}$$

And the operators $A^a$ in (3.9) can be rewritten as:

$$A^1 = -ige^{-ix^*}X, \quad A^2 = -ige^{-ix^*}Y, \quad A^3 = -ige^{-ix^*}Z. \tag{3.12}$$

that is, the $A^a$ operators are, up to a constant factor and a phase factor, just one-qubit quantum logic gates.

It should be noticed that the $A^a$ operators are not Hermitian. This feature is a residual of the dissipative character of the original field theory.

Then, qubits can be processed in the quantum vacuum state of a classical dissipative non-abelian gauge theory, and that decoherence is avoided due to absolute zero temperature of the quantum vacuum. This means that tubulines-qubits of the Penrose-Hameroff model of the quantum mind can take place in this physical model, and moreover they are protected against decoherence.

# 4. Quantum Metalanguage: The non-algorithmic aspect of the mind

The arguments discussed in the previous sections suggest that the non-algorithmic aspect of the mind is hold by a classical, dissipative, non-abelian field theory of the brain. In fact such a theory is not computable, neither classically, nor quantum. The quantum computational aspect is hold by the quantum mechanical vacuum of that theory (in the Hameroff-Penrose model the quantum computational mode should describe the unconscious). The classical-computational mode is obtained after decoherence of superposed quantum states, through interaction with the external world. This mode should correspond to consciousness.

In logical terms, as it was shown in [17], the classical field theory of the brain with hidden quantum nature is a quantum metalanguage (QML), while the quantum mechanics of qubits is the Quantum Object Language (QOL).

QML is made of assertions, linked in a metalinguistic way. The difference with a classical metalanguage is that in QML atomic assertions carry assertion degrees, which are complex numbers, interpreted as probability amplitudes. Also, the QML is equipped with Meta Data, corresponding to the constraint that probabilities sum up to one. The *reflection principle* of basic logic [22] was used to recover QOL from QML. By the reflection principle, all the logical connectives are introduced by solving an equation (called *definitional equation*), which "reflects" meta-linguistic links between assertions into logical connectives between propositions.

The QOL derived from the QML through the *reflection principle*, is made of propositions linked by quantum connectives, like, for instance, the connective "quantum superposition" (the quantum analogous of the classical connective "AND") which is labelled by complex numbers, and is non-commutative.

In the limit of maximal dissipation, when the gauge fields reduce to unitary operators, which process quantum information, we are at the very level of the *reflection principle*: QML is processing QOL whose elements, propositions, are interpreted as quantum states.

It should be noticed that a quantum computer (QC) has a QOL, and its physical theory is QM. Therefore, a QC cannot reach a QML (a non-algorithmic mode of thought) because it is impossible to go from the finite number of degrees of freedom of QM to the infinite ones of FT. That is, a quantum computer will never be able to reach a non-algorithmic mode of thought. This is the difference between a quantum mind and a quantum computer.

In summary, we suggested that the mind has three modes: the non-computational mode (QML), the quantum-computational mode (QOL) describing the Quantum Mind (or unconscious), and the classical-computational one, describing the Classical Mind (or consciousness). The physical description of the first mode is a classical Field Theory, the second one is Quantum Information, the third one is Classical Information.

## Aknowledgements
I am very grateful to E. Pessa and G. Vitiello for useful discussions.